\documentclass[10pt, twoside,a5paper]{article}

\usepackage{german, amsfonts, amsbsy, amssymb, amsmath, times}

\usepackage{graphicx}

\usepackage[utf8]{inputenc}

\usepackage[T2A,T1]{fontenc}

\usepackage{sidecap}

\usepackage{fancyhdr}

\usepackage{natbib}

\makeatletter
\long\def\@makefntext#1{\hangindent 0 cm\noindent
            \hbox{\hspace*{0 cm}$^{\@thefnmark}$}\hspace*{2pt}#1}    
\makeatother

\usepackage[left=2cm,right=2cm,marginparwidth=0cm,
   top=1.1cm,height=196.5 mm,headheight=10pt,headsep=10pt,includehead, includefoot]{geometry}

\usepackage{mathptmx}
\usepackage[scaled=.90]{helvet}
\usepackage{courier}
\renewcommand{\refname}{References}
\begin{document}

\thispagestyle{empty}

 {\centering {\Large \bf Stefan Bergman (1895--1977)\\[4pt]
Notes on His Bio- and Bibliography}\\[10pt]
\renewcommand{\thefootnote}{\fnsymbol{footnote}}
Tilman Sauer\footnote[2]{Email: tsauer@uni-mainz.de. English version of \emph{Stefan Bergmann (1895--1977).
Anmerkungen zur Bio- und Bibliographie}, to appear in: \emph{Beitr\"age zur
Jahrestagung der Fachsektion Geschichte der Mathematik in der DMV},
Ittenbach/K\"onigswinter, 28.--31.Mai 2025, ed.\ H.\,Fischer, M.\,Friedman,
H.\,Heller, R.\,Kaenders, WTM-Verlag, M\"unster.}\\[15pt]}
\renewcommand{\thefootnote}{\arabic{footnote}}

\subsection*{The Early Years}

\noindent
Stefan Bergmann\footnote{In very early documents, the spelling
Stephan occasionally appears; later, the first name was consistently written as
Stefan. After emigrating to the United States, Bergmann began spelling his last
name as Bergman in 1940. For biographical information on Bergman, see 
\cite{SchifferSamelson1979,Skwarczynski1981,Schiffer1981,OConnorRobertson2009}.}
was born on May 5, 1895, in Czenstochau, Silesia---then part of Russia---now 
Cz\k{e}stochowa in Poland.\footnote{There is conflicting information regarding
his year of birth. In the résumés attached to his doctoral and habilitation
files, the year of birth is given as 1895. However, the doctoral file contains
two résumés, one handwritten and one typewritten. It appears that after
submitting the handwritten version, Bergmann was asked to revise his CV, in
which he was required to address, above all, the requirement of fulfillment of
the triennium requirement. In his handwritten CV, Bergmann wrote the
year---presumably by mistake---as 1896. In another handwritten CV from 1933,
attached to the application for support from the Academic Assistance Council 
(note~\ref{note:SPSL}, Bl.\,314f, 320), the year 1897 appears, however, which
is likely an intentional attempt to make himself appear younger in order to
increase his chances of receiving support.} His father, Bronis{\l}aw
Bergmann (1861--1929) was a bank executive, his mother, Tekla Tauba (1861--1934),
née Herc. Bergmann had two younger sisters, Franciszka
(1896--1943?) and Marta (1903-1943?).\footnote{See \cite{Setkowski2020},
entries on Bronis{\l}aw, Adam, Stefan and Marta Bergmann. On the fate of the sisters, see
below, p.\,\pageref{page:bergman-sisters}.} In 1913, he graduated from a state-run eight-year
philological boys' high school and enrolled that same year at the Technical University in
Breslau, where he studied electrical engineering for two
semesters.\footnote{For information on his education and studies, see the doctoral file 
(Phil.Fak.01, Nr. 605) and the habilitation file (Phil.Fak.01, Nr.1245) in the Archives 
of Humboldt University Berlin. A personal file is not listed there,} When World
War I broke out, Bergmann interrupted his studies and worked intermittently in
the workshops of the Berlin tramway company.
In the following three years, 1915–-16, 1916–-17,
and 1917–-18, Bergmann studied at the Vienna University of Technology
and graduated in April 1919 with a degree in electrical engineering. In
1918–-19, he was enrolled as a student at the University of Vienna in the
Faculty of Arts (Mathematics). In the winter semester of 1919–-20,
Bergmann interrupted his studies for health reasons. In the summer semester
of 1920, he enrolled at the University of Berlin and submitted an application
for admission to the doctoral program as early as the beginning of the winter
semester in November 1920.
To demonstrate that he had completed  the required triennium—--that is,
the minimum duration of a three--year course of study---, Bergmann
noted in his curriculum vitae that he had already attended mathematics
lectures during his studies in electrical engineering at the technical
universities in Breslau and Vienna, and indeed ``as an extraordinary auditor
in their entirety.'' As a student in Vienna, Bergmann had already published
two papers on number theory in the \emph{Monatshefte für Mathematik und
Physik} \citep{BergmannS1917Darstellung,BergmannS1918Eigenschaft}.

In his application, Bergmann thanked his doctoral advisor Richard von
Mises (1883-–1953), at whose suggestion and under whose guidance he had
begun work on his dissertation in the summer of 1920. He also thanked
Professors Leon Lichtenstein (1878–-1933), Erhard Schmidt (1876–-1959),
and Issai Schur (1875–-1941). Admission to the doctoral program was granted, and already in May 1921
Bergmann submitted his dissertation, titled ``On the Development of
Harmonic Functions in the Plane and in Space Using Orthogonal Functions''
\citep{BergmannS1922Entwicklung}, for review. Von Mises graded the work as ``valde
laudabile,'' and the second reviewer, Erhard Schmidt, concurred with the
assessment ``durchaus.'' Following an oral defense on June 16, 1921, with
von Mises, Erhard Schmidt, Max Planck (1858–-1947), and Alois Riehl
(1844–-1924) serving as examiners, Bergmann received his doctoral diploma
dated August 13, 1921, with the distinction \emph{magna cum laude}.

His subsequent career began with research work at the Mathematical
Institute; in the summer semester of 1922, he served as an assistant standing
in for Hilda Pollaczek (née Geiringer, 1893–-1973), who gave birth to a
daughter during this time \cite[p.\,211]{Biermann1988}.
According to his own account, Bergmann worked as an engineer from the fall of 1922 to the
summer of 1926, initially in the research department of the AEG turbine
factory in Berlin and later on his own, spending some time abroad as
well.\footnote{As a later CV indicates, this was a position in {\L}odz, Poland.}
In the fall of 1926, he returned to scientific work. From August 1928 to
mid-1930, he conducted research on stability issues in aircraft construction
together with Hans Reissner (1874–-1967) on behalf of the Reich Ministry of
Transport.

\subsection*{Habilitation}

In July 1931, Bergmann submitted his application for admission to the
habilitation process to the Faculty of Philosophy at the University of Berlin.
Attached to the application was a list of 34 publications. The actual
habilitation thesis submitted was a paper titled ``On the Problem of
Ambiguity in Potential Flows with Free Boundaries'' \citep{BergmannS1932Loesungen}.

Bergmann included a note regarding his citizenship in his curriculum
vitae. He stated that he held Polish citizenship (Częstochowa had belonged
to Poland since 1918), but intended to ``take the necessary steps to acquire
Prussian citizenship.'' Bergmann emphasized that
he had lived in Germany or Austria for only ``a single brief interruption'' in
the 18 years since he began his studies in Breslau, and was therefore
confident that Prussian citizenship would be granted to him.

During the faculty’s deliberations on Bergmann’s habilitation, there may
have been a discussion regarding the scope of his venia legendi.\footnote{See
also the much more aggressive debate surrounding Hilda Pollaczek's
habilitation; see \cite[p.\,207f]{Biermann1988}, \cite{SiegmundSchultze1993}.}
Richard von Mises served as the primary reviewer once again, with Ludwig
Bieberbach (1886–1982) acting as the secondary reviewer. Finally, Erhard
Schmidt was brought in as a third reviewer. While von Mises fully
endorsed the habilitation application, Ludwig Bieberbach also argued in his
vote for admission, but expressed the hope ``that, as would be in the interest
of the department, which has few faculty members in the field of applied
mathematics, Mr. Bergmann might shift the focus of his teaching to applied
mathematics and not allow himself to be tempted, by the granting of an
unlimited venia, to turn his back on applications in his teaching.'' Schmidt,
too, without providing his own report, spoke in favor of admission.

Of the three topics for the colloquium: 1.\ Numerical methods for
calculating mapping functions, particularly for polygonal regions,
2.\ On shear buckling of plates, 3.\ The application of orthogonal functions in
the theory of functions of two complex variables, the third was selected, and
of the three topics for the inaugural lecture: 1.\ Applications of complex analysis
methods in the theory of elasticity, 2.\ Numerical methods for solving
variational problems, 3.\ Problems and tasks in the theory of functions of two
complex variables, the second was selected. The habilitation colloquium
took place on March 2, 1932; participants in the discussion included von
Mises, who, as dean, also chaired the discussion, as well as Bieberbach and
Schmidt. As a result, the \emph{venia legendi} was granted for the field
``Mathematics,'' with the words  ``pure and applied'' explicitly added by hand
to the minutes.

Bergmann, however, did not enjoy the teaching authority he had earned
for long. During the winter semester, he gave a two-hour lecture (Tuesdays
and Thursdays, 9–10 a.m., privately, i.e., for a fee) on his specialty,
``Analytical Functions of Two Variables.''\footnote{See
\emph{Friedrich-Wilhelms-Universität zu Berlin. Vorlesungsverzeichnis.
Wintersemester 1933.} 
  Preussische Druckerei und Verlagsaktiengesellschaft Berlin, p. 58.}
Interestingly, this was precisely
a topic from the field of pure mathematics, and one in which Bieberbach
himself was an expert. In addition, Bergmann taught
the ``Seminar on Applied Mathematics (Mechanics)''
at the Institute for Applied Mathematics,
together with Richard von Mises and Hilda Pollaczek.

For the summer semester of 1933, Bergmann announced, in addition to
the seminar on applied mathematics with von Mises, a two-hour lecture on
``Theory of Elasticity'' (Sundays, 9-–11 a.m., also privately).\footnote{See
\emph{Friedrich-Wilhelms-Universität zu Berlin. Vorlesungsverzeichnis.
Sommersemester 1933.} Preussische Druckerei und Verlagsaktiengesellschaft
Berlin, p.\,56f.}
However, this never came to pass.\footnote{For the following, see the file
``Durchf\"uhrung des Berufsbeamtengesetzes hinsichtlich der Professoren und Privatdozenten in der Philosophischen Fakult\"at Berlin'' (``Implementation of the Civil Service Act with regard to
professors and private lecturers in the Faculty of Philosophy at the University
of Berlin,'' \emph{Geheimes Preussisches Staatsarchiv}, Rep.76 Va Sect.2 Tit.IV, pp.
548–-555.  \label{note:GehPrSta}}

On April 7, 1933, the Nazis enacted the so-called Law
for the Restoration of the Professional Civil Service. That same month, the
university’s administrative director sent out questionnaires asking, among
other things, about the racial affiliation of grandparents and participation in
the World War. The form was to be returned by April 29, but Bergmann
wrote to the administrative director on April 27 that he had ``not yet
answered the form because I have not been able to ascertain several facts to
date.'' Furthermore, he pointed out that he was a Polish citizen and that the
questionnaire therefore ``probably did not apply to him.'' Additional notes
on the letter mention that he was born on May 5, 1896[sic] in Częstochowa,
had received his habilitation on March 2, 1932, and was of Jewish descent.
It is also noted that Bergmann
``had not read S.S.33,'' i.e.\ in the summer semester 1933. Finally, in a letter
dated September 24, 1933, he was
informed that, pursuant to § 3 of the law—that is, due to his non-Aryan
ancestry—his teaching license at the University of Berlin had been
revoked.\footnote{See also \cite[p.\,13]{Asen1955}, \cite[p.\,237, 348]{Biermann1988}.}\\

\subsection*{The Odyssey of an Emigrant}

It is unclear where Bergmann was at this time. A letter dated April 27, 1933,
addressed to his Berlin address (Flensburger Str.\,8) was forwarded to a post
office box in Częstochowa.

Skwarczy\'nski (\citeyear[p.\,193]{Skwarczynski1981})
reports that Bergmann sold electrical
appliances for several months in a store financed by his family, but that the
business eventually went bankrupt.

On July 2, 1933, Bergmann applied for financial support from the British
\emph{Academic Assistance Council}.\footnote{See
the file ``MS. S.P.S.L 277-5 Bergmann, S,'' Bodleian Library, Oxford, UK. The AAC
was renamed the Society for the Protection of Science and Learning (SPSL) in 1936 and has
been active since 1999 as the Council for Assisting Refugee Academics (CARA).
\label{note:SPSL}}
The AAC had been established in April
1933 by William Beveridge (1879-–1963) in response to the dismissals of
Jewish scholars in Germany. A founding document dated May 22, 1933,
lists two objectives of the organization’s activities: first, 
to raise public funds to support scholars in need and, secondly, to act as a
placement agency assisting emigrants in their job search. Although based in
England, the AAC also sought contact with similar initiatives in other
countries, such as the Emergency Association of German Scientists Abroad,
founded at the same time in Zurich. On October 3, 1933, the AAC held a
large gathering at London’s Royal Albert Hall to draw attention to its goals
and raise funds. The prominent speaker at this gathering was Albert
Einstein, who delivered his widely acclaimed address ``Science and
Civilization'' there.

Bergmann attached an academic résumé (pp.\,317ff., 320) and a list of
publications now comprising 46 titles (pp.\,318ff.) to his application for
support from the AAC. Bergmann also referred to his financial situation. He
was unmarried, his father had passed away, his mother’s financial situation
was ``quite difficult,'' and his two sisters were unmarried. He possessed no
assets and lived on loans from his relatives. ``My lecturing activities in
Berlin and, in connection with that, the possibility of earning an occasional
income through scholarly work are currently cut off for me.''

The AAC and presumably Bergmann himself requested statements from
a number of scholars, and Bergmann’s AAC application file includes a
number of very positive letters of recommendation. Richard von Mises, who
was himself in the process of moving to the University of Istanbul, wrote
very positively, as did Jacques Hadamard (1865–-1963) from Paris. From
Great Britain, there were letters of recommendation from E.T.\,Whittaker
(1873–-1956) of Edinburgh, from S.\,Brodetsky (1888–1954) of Leeds, from
G.H.\,Hardy (1877–-1947) from Cambridge, and from G.B.\,Jeffery (1891–-1957) 
from London (pp.\,321–326). However, there is no evidence that
Bergmann’s application was successful. A letter from the AAC to the Joint
Foreign Committee, to which Bergmann had apparently also applied for
support, indicates that his Polish citizenship was a disadvantage, as the
AAC advised nationals of other countries to seek refuge in their home
countries. Furthermore, they reportedly had insufficient funds. Finally, a
rather half-hearted recommendation from
R.V.\,Southwell (1888–-1970) from Oxford (p.\,327) likely did not contribute
to a positive decision.

The correspondence in the S.P.S.L.\,file indicates that Bergmann stayed in
Paris, at least temporarily, in 1933 and 1934. Bergmann also published 
a French-language paper in the Comptes rendus, presented by E.\,Cartan at
the meeting of July 23, 1934.

From the fall of 1934, Bergmann was a professor at Tomsk State
University. The Novosibirsk newspaper \emph{Sovetskaya Sibir} reported on this in
its September 20, 1934, edition in a brief note titled {\fontencoding{T2A}\fontfamily{ptm}\selectfontПриезд немецких профессо\-ров}.`` A German translation of the newspaper report was immediately sent by the
German Consulate in Novosibirsk to the Foreign Office in Berlin and
reads:\footnote{See footnote \ref{note:GehPrSta}, pp.\,1183--1184.}
\begin{quote}
	Arrival of German Professors. Two of the most prominent
foreign specialists, Professors Bergmann and Netter, who were
expelled from Germany by the government of the fascist
dictatorship, have arrived to work at the Scientific Research
Institute for Mathematics and Mechanics at Tomsk State
University.
Professor Bergmann is a specialist in complex analysis, one of
the most prominent specialists in Europe. He works in the field
of applied mathematics at Professor Mises’s scientific research
institute at the University of Berlin. Bergmann has published
approximately 40 papers.
\end{quote}
Bergmann presumably remained in Tomsk for two years, until the summer
of 1936. According to a note in Bergmann’s S.P.S.L. file regarding a
communication from Hans Georg Baerwald (1904–1987) in London dated
December 6, 1937, Bergmann was at the Academy of Sciences in Tbilisi at
that time.\footnote{The mathematics institute, now named after Andrea Razmadze, was part of Tbilisi State
University, founded in 1918. “On October 1, 1935, [...] the mathematics and mechanics section
of the [...] institute was transformed into a mathematical research institute under the auspices of
the Georgian Branch of the USSR Academy of Sciences. The institute was incorporated into
the Georgian Academy of Sciences after the latter was founded in February 1941. In 1944, the
institute was named after A. Razmadze.” ftp.rmi.ge/emg/about.htm, last accessed in January
2026.}
The note reads ``Left Tomsk, where unpopular.'' However, in a letter from Baerwald to the
long-time secretary of the S.P.S.L., Esther Simpson (1903–-1996), dated
January 7, 1938, Baerwald reports that he had heard in the USSR that
Bergmann had left the USSR for his summer vacation in 1937 but had not
returned. In fact, Pinl and Furtmüller (\citeyear[p.\,156]{Pinl1973}) also write that
Bergmann
``was forced to leave the USSR as well,'' and his biographers \cite{SchifferSamelson1979} 
also report that his position in the USSR had become
precarious in the wake of Stalin’s purges.

In January 1938, Esther Simpson noted in the journal \emph{Nature} a reference
to a lecture Bergmann had given at the Académie des Sciences in Paris during
the session of December 13, 1937.\footnote{\emph{Nature} 141(29.\,Jan.\,1938), p.\,216.
}
She subsequently wrote to Emil Julius
Gumbel (1891–1966) in Lyon, asking whether this meant that Bergmann
had lost his position in the USSR. Gumbel replied that Bergmann was only
temporarily on leave from Tbilisi and was staying in Paris to complete a
book that was to be published in
French.\footnote{\citep{BergmannS1947Fonctions,BergmannS1948Fonction}, the
manuscripts of which
had already been submitted in early
1939 but could not be published until later due to the war; see the note in
\cite[p.\,6]{BergmannS1948Fonction}.}

Even before the outbreak of the war, however, Bergmann had already left France
again and traveled to the United States, where he arrived in New York City on
May 16, 1939.\footnote{For the following, see the file \grqq{}Bergman, Stefan,
cuid29421, Box 30, Folder 5, Office of the President, records of Karl Taylor
Compton and James Rhyne Killian, AC-0004.\grqq{} Massachusetts Institute of
Technology, Libraries.}
As indicated in a memo dated September 21, 1939, Bergmann
initially accepted an invitation from the Harvard Engineering School on the
recommendation of Richard von Mises, but soon joined the mathematics department
at MIT. On the recommendation of Norbert Wiener (1894–1964), Bergmann was then
hired effective December 1 as a lecturer in mathematics with an annual salary
of \$900. The appointment was made primarily with a view to extending Bergmann’s
visitor visa. This stipend was described as “nominal, since it has been
indicated that Dr. Bergmann has some private resources.”

Bergmann’s next stop was Yeshiva College in New York City. There,
Bergmann received a two-year contract as an \emph{instructor in mathematics} for
the academic years 1940–-41 and 1941–-42 with an annual salary of 
\$\,1,000.\footnote{Letter of appointment from President Revel to Bergmann dated April 22, 1940, Yeshiva
University Library, Special Collections and Hebraica-Judaica, Correspondence Revel, Box 4,
Folder 21-3-4.}
His appointment took place as part of a reform of the college and an
expansion of its graduate school by President Bernard Revel.

Bergmann then spent the following years, 1942–46, at Brown University
in Providence, Rhode Island.\footnote{A brief and not very accurate index card
(year of birth: 1899!) lists only one appointment as
\emph{Visiting Lecturer} for the year 1941. Brown University Library, Special Collections,
Biographical Files, Box 47, Stefan Bergmann 1-S.}
The university yearbooks for 1942 and 1943
list Bergmann as a \emph{visiting lecturer} in the mathematics faculty.\footnote{\emph{Liber brunensis} 1942, p.\,205;
No yearbooks were published in 1944 and 1945.}
Other émigrés in the mathematics faculty at the time included, in addition to Otto
Neugebauer (1899–-1990), Richard von Mises (for a time), Georg Pólya
(1887–-1985), and Léon Brillouin (1889–-1969). Bergmann’s work at Brown
University took place as part of a program launched in the summer of 1942
that aimed to prepare technically trained applied mathematicians for
deployment in the war industry. The \emph{Graduate School of Advanced
Instruction and Research in Mechanics} focused on aeronautics,
aerodynamics, acoustics, and submarine communications, with courses in
``elasticity, fluid dynamics, advanced dynamics, and hydrodynamic theory
of propellers.''\footnote{J.\,Henderson,
\glqq{}Mechanics Theories Emphasized in Expanding Graduate School.\grqq{} \emph{Brown Daily
Herald}, July 15, 1942, p.\,4.}
In an interview, Bergmann described his role in training engineers for submarine warfare:
``Applied mathematics coordinates the work of the
pure mathematician and the engineer in the task of perfecting American
submarines. One of the chief purposes of the mechanics summer school is to
train young men to apply the theory of pure mathematics to engineering
fields through this medium.''\footnote{S.P.\,Culviner, ``New War
Mechanics Courses Aim At Perfection of Sub Chasers.'' \emph{Brown Daily
Herald}, August 7, 1942, p.\,1.}

\label{page:bergman-sisters}
During this time, Bergmann lost the remaining members of his family.
His father had died in 1929, and his mother in 1934 in Cze˛stochowa. On
April 5, 1943, Bergmann wrote to Esther Simpson regarding his two sisters,
Franciska (born February 17, 1896) and Marta (born January 31, 1903), as
well as their husbands, his brother-in-law Władysław Haltrecht from Łódź,
and asked for help (p. 368). At that time, the three were living in the
Warsaw Ghetto (23 Grzybowska). Details about their subsequent fate are
unknown, but in January 1969, the New York Times reported on a
fundraising drive ``plight of the needy,'' to which Bergman also contributed, ``in memory of
Anka Bergman and Marta Haltrecht, `murdered in Hitler’s concentration
camp.' ''\footnote{\emph{New York Times}, January 10, 1969, p.\,52.}
Skwarczy{\'n}ski (\citeyear[p.\,196]{Skwarczynski1981})
reports that during his frequent
visits to Poland after the war, Bergman regularly visited the graves of his
sisters, who had been murdered by the Nazis.

Bergmann dedicated his first two books, which he had written in Paris
before emigrating to the United States but which were not published until
after the war
\citep{BergmannS1947Fonctions,BergmannS1948Fonction}, to his father Bronis{\l}aw.
He dedicated
a book published in 1950 on ``The Kernel Function and Conformal Mapping''
``to the memory of my sisters Franciszka Anka Bergman and Marta
Haltrecht'' \citep{BergmanS1950Function}. He dedicated another book in 1963 to his
mother: ``To the memory of Tekla Bergman née Hertz''
\citep{BergmanS1961Operators}.

\subsection*{The Postwar Years}

After the war, Bergman first went to Cambridge, MA. From the academic
years 1946/47 through 1950/51, Bergman is listed in the annual Harvard
University Directory of University Officers and Students as a \emph{Research
Lecturer in Aeronautical Engineering}. He taught at the \emph{Graduate School of
Engineering} during the academic years 1947/48, 1948/49, and
1949/50.\footnote{Official Register of Harvard University XLVII, May 16, 1950,
No.\,12, pp.\,366, 368; XLIX, April 30, 1952, No.\,10, pp.\,406, 408i; LI, April 2,
1954, No.\,6, pp.\,82.}
specifically in the Department of Engineering on ``Partial Differential
Equations, with Applications to Fluid Dynamics.'' It can be assumed that a
significant portion of the funding for Bergman’s work during these years
came from external sources. In any case, Bergman published some of his
research findings as technical reports of the \emph{National Advisory Council for
Aeronautics} (NACA), the predecessor institution of today’s NASA. His publications
cite a wide range of funding agencies, such as the Office of Naval Research
(ONR), among others.

In 1951, Bergman was elected to the American Academy of Arts and
Sciences.\footnote{www.amacad.org, last accessed in January 2026.}
While still in Cambridge, in 1950, Bergman married Adele
Latzer, née Adlersberg.\footnote{Details on Adele Adlersberger, in
\emph{Biographisches Handbuch der deutschsprachigen Emigration nach 1933},
Vol.\,2. 1983 (565).} The couple had no children.
It was also
during this time that Bergman began an intensive collaboration with the
mathematician Menachem Schiffer (1911–1997), with whom he published a
whole series of joint papers during those years. In the fall of 1952, Schiffer
moved to the West Coast to Stanford, and according to Krantz (\citeyear[p.\,33]{Kantz1990}),
Schiffer was instrumental in helping Bergman secure a position there as
well.\footnote{Stanford University Library, Special
Collections, holds Bergmann’s correspondence
collection under call number SC0404, which consists largely of the scholarly correspondence
between Bergmann and Hilda Geiringer (née Pollaczek-Geiringer, widow of Geiringer-von
Mises).}

Bergman also moved to Stanford University in 1952 and remained in
Palo Alto, California, until his death in 1977, where, as his biographers
unanimously report, he spent his happiest years.\footnote{Impressions of Bergmann’s 
personality are described by \cite{Skwarczynski1981,Kantz1990,Davis1997}.}

After his wife’s death, a foundation was established from her estate,
which funded the \emph{Stefan Bergman Prize} from 1989 to 2023. Prize recipients
were selected by the \emph{American Mathematical Society}, and the prize money
most recently amounted to \$24,000. Since 2023, the prize has been
converted into a \emph{Stefan Bergman Fellowship} for ``early-career
mathematicians.''

\vspace{-0.5 cm}

\setlength{\bibsep}{0.0pt}

\vspace{-5mm}

\nocite{*}
\bibliographystyle{apalike}
\section*{Additions to Bergman's Bibliography\footnote{Skwarczynski’s bibliography
of Bergmann’s writings  (\citeyear{Skwarczynski1981}) comprises 160 titles and is
quite comprehensive for the period following Bergmann’s emigration to the United States, but
incomplete for his early years. The basis for these additions to Skwarczynski’s bibliography
was primarily the lists attached to Bergmann’s habilitation file, the list of works in his SPSL
file, as well as the entries in Poggendorff,  \emph{Biographisch-literarisches
Handwörterbuch ...}, Vol.\,VI: 1923--1931, all of which are based on
information provided by Bergmann himself.}}
\vskip -1cm
\renewcommand\refname{}
\bibliography{bergmanbibadd}

\end{document}